\documentclass[12pt]{article}

\usepackage{graphicx}% Include figure files
\usepackage{amsmath,amssymb}
\usepackage{hepunits}
\newcommand{\beq}{\begin{equation}}
\newcommand{\eeq}{\end{equation}}
\newcommand{\beqa}{\begin{eqnarray}}
\newcommand{\eeqa}{\end{eqnarray}}
\newcommand{\beqar}{\begin{eqnarray*}}
\newcommand{\eeqar}{\end{eqnarray*}}

\begin{document}
\thispagestyle{empty}

\hfill{\sc UG-FT-286/11}

\vspace*{-2mm}
\hfill{\sc CAFPE-156/11}

\vspace{32pt}
\begin{center}
{\textbf{\Large Gluon excitations in $t\bar t$ production}}

{\textbf{\Large at hadron colliders}}

\vspace{40pt}

R.~Barcel\'o, A.~Carmona, M.~Masip, J.~Santiago
\vspace{12pt}

\textit{
CAFPE and Departamento de F{\'\i}sica Te\'orica y del Cosmos}\\ 
\textit{Universidad de Granada, E-18071, Granada, Spain}\\
\vspace{16pt}
\texttt{rbarcelo@ugr.es, adrian@ugr.es, masip@ugr.es, jsantiago@ugr.es}
\end{center}

\vspace{40pt}

\date{\today}% It is always \today, today,

\begin{abstract}

We argue that a relatively light massive gluon with mass $\lesssim 1$
TeV, small purely axial couplings to light quarks and sizable vector and
axial couplings to the top quark 
can reproduce the large forward-backward asymmetry observed at the
Tevatron without conflicting with the $t\bar{t}$ and the
dijet invariant mass
distributions measured at the Tevatron and the LHC.
We show that realistic Higgsless models with warped extra dimensions 
naturally fulfil all the necessary ingredients to realize this
scenario. While current data is unable to discover or exclude these
heavy gluons with masses $\approx 850$ GeV, they should be observed at
the (7 TeV) LHC with a luminosity $\gtrsim 300$ pb$^{-1}$.

%\pacs{}
%PACS, the Physics and Astronomy Classification Scheme.
%\keywords{}

\end{abstract}

\newpage

\section{Introduction}

The standard model (SM) does not {\it explain} the difference between the
electroweak (EW)  and the Planck scales ($M_{EW}^2\approx
10^{-32}M_{Planck}^2$).
The large value of the top mass makes it plausible that any new
physics responsible for this difference would
show up in top physics. Thus a detailed study of top properties is one
of the main goals of the Tevatron and the LHC. In fact, Tevatron
experiments have already observed a significant anomaly in the
forward-backward asymmetry $A^{t\bar{t}}$ in $t\bar{t}$
production~\cite{AFB1,AFB2,AFB3}, an anomaly that
is not present in the total cross section (\textit{i.e.}
integrated over all angles). It is not easy to explain this anomaly
with
current Tevatron and LHC data on dijets and top-pair production. 
New heavy color
octect gauge bosons with axial couplings to the SM quarks, axigluons,
have been proposed as possible
candidates~\cite{Ferrario:2009bz,Antunano:2007da,Burdman:2010gr} 
(see also~\cite{Degrande:2010kt} 
for a model independent discussion). The
reason, that we review below, is that axial couplings contribute
maximally to the asymmetry but cancel to leading order in the total
cross section.

In this article we point out a region of parameter space of these color
octects that has been overlooked in the past.~\footnote{A related
  discussion, with emphasis on fourth-generation quark production in
  the context of models of strong EW symmetry breaking can be found
  in~\cite{Burdman:2010gr}.} It
corresponds to relatively light $m_G \lesssim 1$ TeV gluons with small
axial couplings to the light quarks and order one axial and vector
couplings to the top quark. As we will show, the axial nature of the
couplings to light quarks and the mass of the new gluon are enough to
hide it from Tevatron data on the total cross section and the
$t\bar{t}$ invariant mass distribution ($M_{t\bar{t}}$) 
while agreeing with the observed asymmetry. At the LHC the
small couplings to the light quarks (and the zero coupling to gluons) makes
the heavy gluon invisible in dijet data, whereas the luminosity of
the 2010 LHC run is not enough to make it visible in $t\bar{t}$
data. It should however show up clearly in the data to be collected
during the 2011 run.

Interestingly enough,
Higgsless models with warped extra
dimensions~\cite{Csaki:2003dt,Cacciapaglia:2006gp}   
naturally realize the scenario we have just discussed. In these
models there are massive copies, the Kaluza-Klein (KK) excitations, 
of the SM gauge bosons, including the gluon, with masses bound from
below by EW precision tests to be $M_G \gtrsim 0.7$ TeV and from above
from the fact that the KK excitations of the EW gauge bosons
have to unitarize longitudinal gauge boson scattering, $m_G \lesssim
1$ TeV (the masses of the KK excitations of the gluons and the
EW bosons are of the same order). At the same time it is
natural that the left and right handed components of the SM fermions are
localized at different points of the extra dimension, which means that
their couplings to the KK gluons are in general different.
Therefore, KK gluons 
will have both vector and  axial-vector couplings to fermions:
\beq
g_V^f={g^f_R+g^f_L\over 2}\;,\;\;\;g_A^f={g^f_R-g^f_L\over 2}\;.
\label{gf}
\eeq
As we will see below, 
constraints on these models from EW precision data 
tend to require the couplings of the light quarks
to the KK gluon to be small and mainly axial, as it is also 
preferred by top data.  
Such axigluons have another {\it unusual} 
feature, namely, they do not decay into massless gluons. This can
be easily understood from the orthogonality of their wave functions:
the overlap between an initial massive mode 
and the two final (delocalized) gluons adds always to zero.
The KK excitations are then far from being 
massive replicas of the standard zero mode, as often assumed in
collider searches. Here we study under what conditions
they are consistent with the data on $t\bar t$ 
production at the Tevatron and with LHC data.

The outline of the paper is as follows. In the
section~\ref{axigluonsatTeV} 
we review
the effect of new heavy gluons with vector and axial couplings on
Tevatron data. In section~\ref{higgsless} we describe the
relevant features of realistic
Higgsless models and how they naturally realize new axigluons
compatible with Tevatron data. Section~\ref{LHC} is devoted to the
implication of realistic models at the LHC and finally we conclude in
section~\ref{conclusions}.

\section{Vector and axial-vector gluons at the Tevatron
\label{axigluonsatTeV}}
Before considering a more motivated model, we review 
the impact that a massive  gluon $G$ may have on $t\bar t$ physics in
the simplest cases (for a related discussion
see~\cite{Ferrario:2009bz,Antunano:2007da,Burdman:2010gr}).  
We will focus on 
$M_{t\bar t}$ and $A^{t\bar t}$,
two observables that have been measured at the Tevatron with an
integrated luminosity up to 5.3 fb$^{-1}$. We use the
parton level asymmetry in the $t\bar{t}$ rest frame that has been
measured to be~\cite{AFB3}
\begin{equation}
A^{t\bar{t}}=0.474\pm 0.114. \label{AFB:data}
\end{equation}

We consider two
different options according to the coupling of $G$ to the light
quarks:
\begin{itemize}
\item Coupling to vector currents ({\it case V}):
\beq
g_V^q=g^q_R=g^q_L\,;\;\;\;g_A^q=0\;.
\label{g1}
\eeq
\item Coupling to axial currents ({\it case A}):
\beq
g_A^q=g^q_R=-g^q_L\,;\;\;\;g_V^q=0\;.
\label{g2}
\eeq
\end{itemize}
For the top quark we will simply assume
\beq
g^t_R\ge g^t_L>0\;.
\label{gt}
\eeq

\begin{figure}[h]
\begin{center}
\begin{tabular}{cc}
\includegraphics[width=0.5\linewidth]{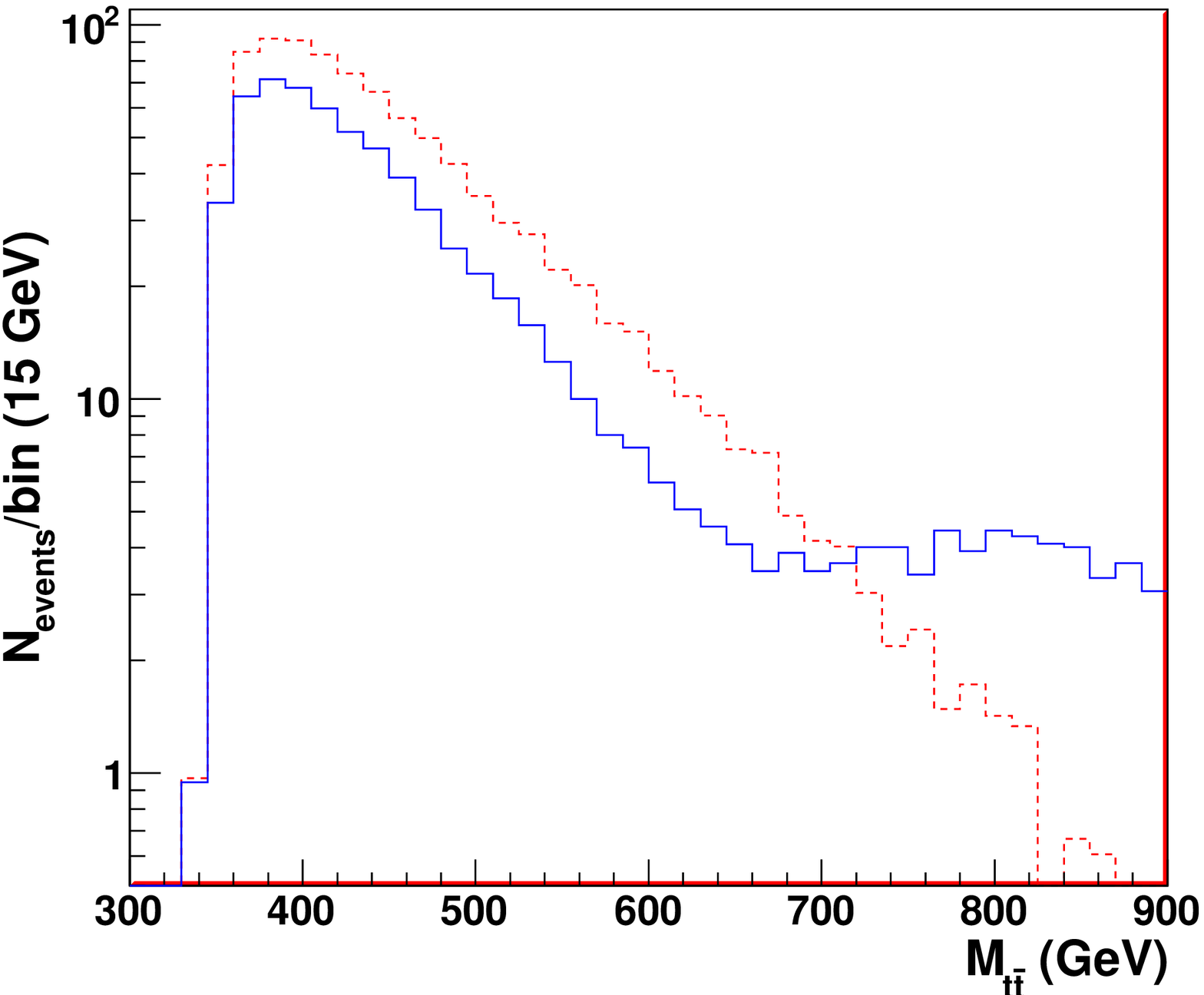} & 
\includegraphics[width=0.5\linewidth]{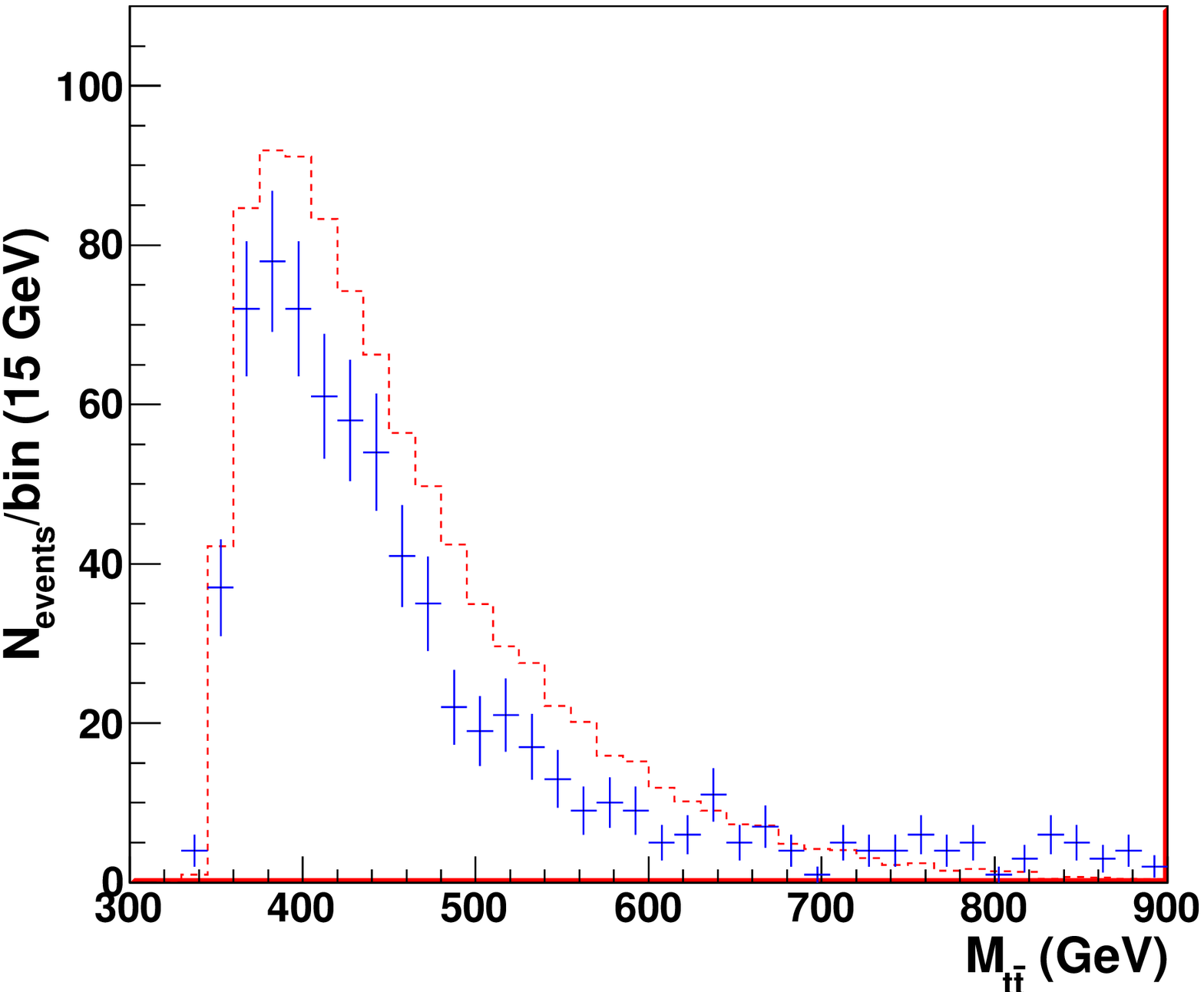} 
\end{tabular}
\end{center}
\caption{$M_{t\bar t}$ distribution at the Tevatron 
in the SM (dashes) and in model $V$ (solid) 
for a luminosity of 5.3 fb$^{-1}$ and $g_V^q=+ 0.2$.
On the left we plot the average number of events expected in each 
case, and on the right a particular Montecarlo simulation. 
The errors shown are statistical only.
}
\label{fig1}
\end{figure}
We have implemented the model in MADGRAPH/MADEVENT
v4~\cite{Alwall:2007st}, used PYTHIA~\cite{pythia} for hadronization
and showering and PGS4~\cite{PGS4} for detector simulation. 
In Fig.~1 we plot $M_{t\bar t}$ distribution for {\it case V}
with $(g_V^q=0.2\,g,g_A^q=0)$,  
$(g_R^t=6\,g,g_L^t=0.2\,g)$ and a mass $M_G=850$ GeV. 
For these couplings the gluon width is
$\Gamma_G\approx 0.32 M_G$. 
We have taken an integrated luminosity of 5.3 fb$^{-1}$
and the cuts/acceptances described in~\cite{AFB3} (we have normalized
our samples so that our SM prediction 
agrees with the background-subtracted data of~\cite{AFB3}). 
The $682$ semileptonic $t\bar t$ pairs given by this model 
(see Fig.~1--{\it left}) result from the destructive 
interference of the standard [$\approx g^2/\hat s$]
and the massive-gluon [$\approx 0.2g\cdot 6g/(-M^2)$] amplitudes.
We obtain a 30\% reduction for 
$M_{t\bar t}<M_G-\Gamma_G$ and an excess
at higher invariant masses with respect to the SM. The distribution does not
show a clear peak, but the change in the {\it slope} at 
$M_{t\bar t}\approx 650$ GeV
would have been apparent in the data. Taking the opposite sign
for the light-quark vector coupling $(g_V^q=-0.2\,g,g_A^q=0)$ the 
situation is similar, although the interference is now constructive
at low values of $M_{t\bar t}$.

In these models the forward-backward asymmetry will 
appear only at next-to-leading order, since 
$A_G^{t\bar t}\propto -g_A^qg_A^t=0$ (see for
example~\cite{Ferrario:2009bz}).  
In particular, 
the  interference of 
the tree-level and  the one-loop box amplitudes will provide the
standard contribution, of order $A_{NLO}^{t\bar t} \approx 0.09$
at high invariant masses as estimated in~\cite{AFB3} using
MCFM~\cite{Campbell:1999ah}.  
An analogous interference between the massive gluon 
and the box diagrams will also contribute to 
the asymmetry.
At $M_{t\bar t}\ll M_G$ we estimate (see also~\cite{Bauer:2010iq})
\beq
A_{V-NLO}^{t\bar t} \approx A_{NLO}^{t\bar t} \times {M_{t\bar t}^2\over -M_G^2}
{g_V^q g_V^t\over g^2}\,,
\eeq
implying an additional contribution
of order $A_{V-NLO}^{t\bar t} \approx \mp 0.04$ for $g^q_V=\pm 0.2\,g$.
Therefore, 
the total value seems in this {\it case V}
very far (over 3 $\sigma$) from 
the asymmetry deduced from
the Tevatron data.

\begin{figure}
\begin{center}
\begin{tabular}{cc}
\includegraphics[width=0.5\linewidth]{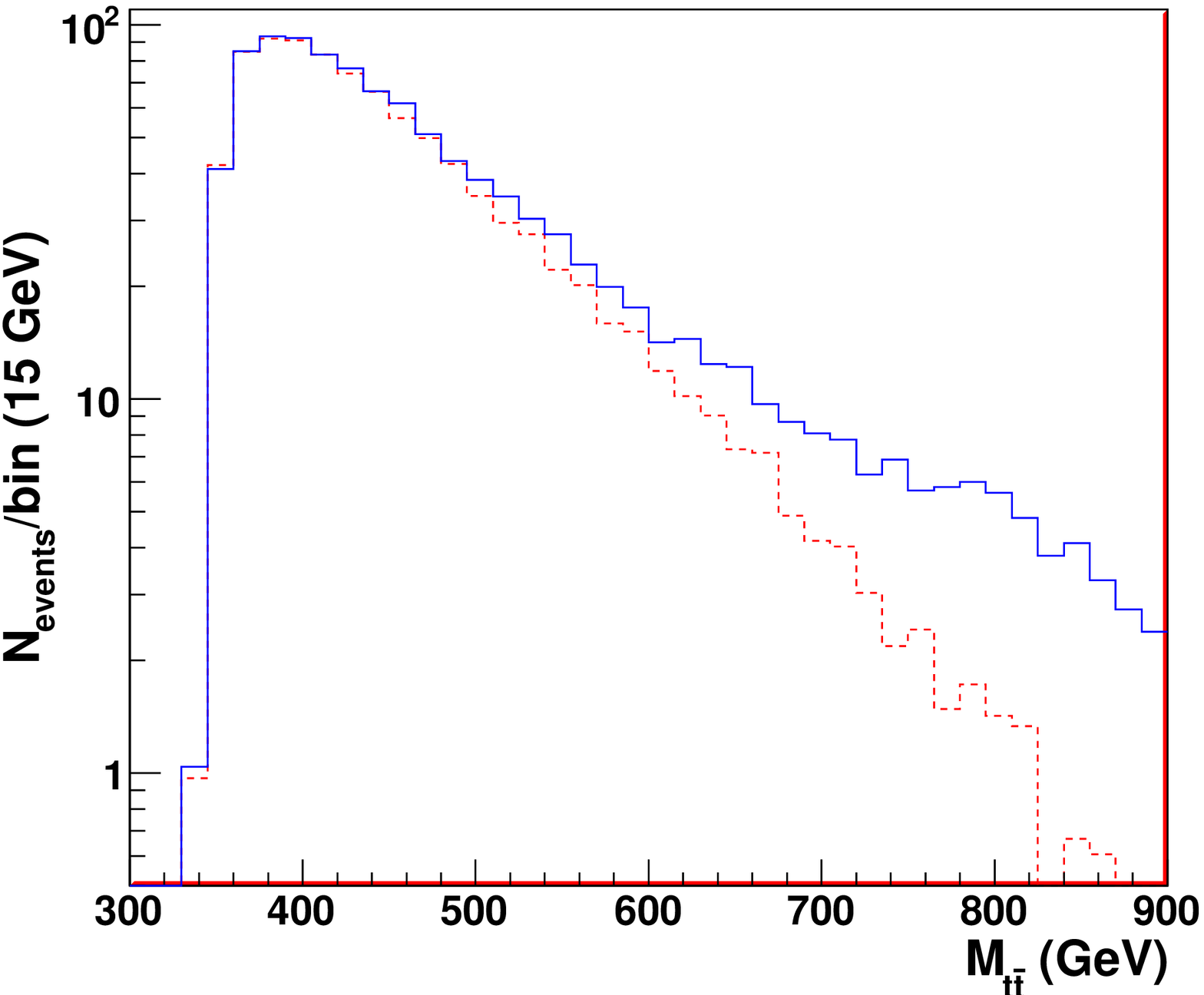} & 
\includegraphics[width=0.5\linewidth]{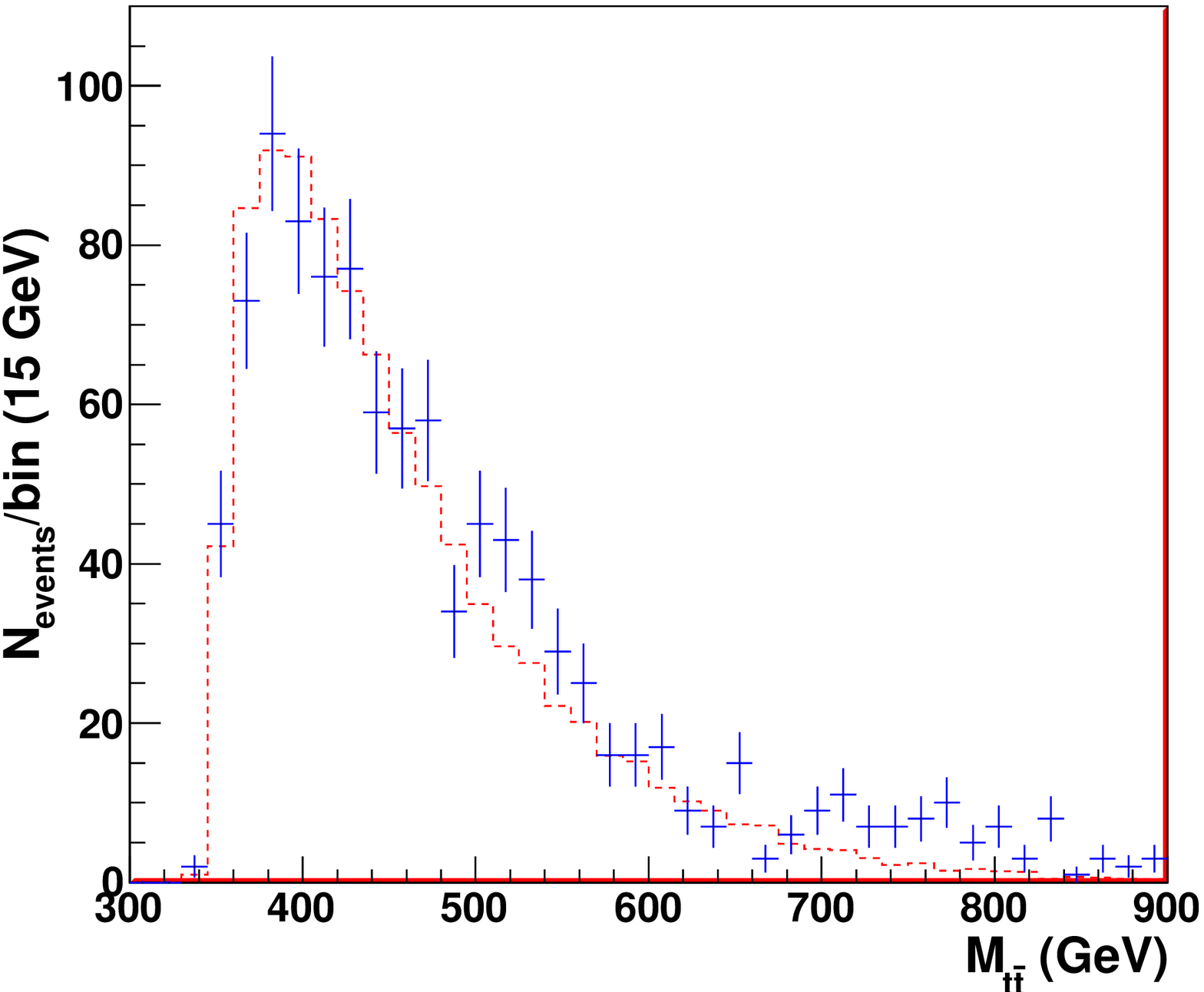} 
\end{tabular}
\end{center}
\caption{$M_{t\bar t}$ distribution at the Tevatron 
in the SM (dashes) and in model $A$ (solid) 
for a luminosity of 5.3 fb$^{-1}$ and $g_A=-0.2$.
On the left we plot the average number of events expected in each 
case, and on the right a particular Montecarlo simulation.
The errors shown are statistical only.}
\label{fig2}
\end{figure}
{\it Case A}, with a purely axial-vector coupling to the light
quarks, is completely
different. Both $q_L\bar q_L\to t\bar t$ and $q_R\bar q_R\to t\bar t$
parton-level cross sections will 
have large contributions from the interference. 
However, since their couplings
are opposite ($g_L^q=-g_R^q$), it will be constructive
in the first process and destructive in the 
second one, and both effects tend to cancel each other.
Up to invariant masses $M_{t\bar t}\approx M_G-\Gamma_G$ where the
resonant contribution becomes important, the number of $t\bar t$
events and their $M_{t\bar t}$ distribution will be very close to
the one in the standard model. Note that the top couplings do not need
to be purely axial for this to happen. The region around the peak 
will be hidden by the low statistics if $M_G$ is large
enough. In Fig.~2 we plot {\it case A} with
$(g_A^q=-0.2\,g,g_V^q=0)$,  
$(g_R^t=6\,g,g_L^t=0.2\,g)$, $M_G=850$ GeV and  
$\Gamma_G=0.32 M_G$ GeV. After cuts we obtain
1042 $t\bar t$ pairs, a number only 12\% higher than
the one expected in the standard model. At
$M_{t\bar t} \approx 600$ GeV the distribution exhibits
a change in the 
slope, but the region where the
differences are important (around 750 GeV) is of 
little statistical significance (see a 
particular Montecarlo simulation in Fig.~2--{\it right}).
Notice that in this model the peak 
at $M_{t\bar t} = 850\pm 272$ GeV is
practically nonexistent, so it would
be challenging to exclude it at the Tevatron even with 
an increased luminosity.

In contrast to the case with vector couplings to the light quarks, 
$A_{G}^{t\bar t}$ is in {\it case A} large: the total
number of events does not change, but there is a large
forward excess that coincides with the backward deficit. 
In the $t\bar t$ rest frame we obtain
\beq
A_{G}^{t\bar t} \approx \left\{
\begin{array}{l l} 
\displaystyle 0.07
& M_{t\bar t}<450\;{\rm GeV}\,; \\
0.20
& M_{t\bar t}>450\;{\rm GeV}
\;.
\end{array} \right. 
\eeq

Therefore, {\it case A} provides a promising framework for model
building. Such a light axigluon could in principle be strongly
constrained by flavor data. We show in the next section that 
Higgsless models with warped extra
dimensions naturally realize the framework we have
just described here. In such models one can implement
flavor symmetries that keep these flavor constraints under
control~\cite{Csaki:2009bb}. Also note that the couplings
of the $(t_L\; b_L)$ doublet with
the axigluon do not need to be too large in order to generate 
a sizable $A^{t\bar t}$ thus further reducing constraints from B
physics~\cite{Chivukula:2010fk}.

\section{Axigluons in a realistic Higgsless model\label{higgsless}}

We consider the realistic warped Higgsless model proposed
in~\cite{Cacciapaglia:2006gp} in which the EW symmetry is broken via
boundary conditions. This can be understood as
a limit with $\langle H\rangle\to \infty$ that forces
the $W,Z$ wave functions to vanish at the IR brane
keeping their masses finite,  
while the physical (4dim) Higgs decouples. The $Z$ and $W$ bosons
become then {\it anomalous} KK modes, much lighter than 
higher excitations and with a flatter wave function along the 
extra dimension. 
The model and its EW constraints are described in some detail in the
appendix, here we just emphasize the most relevant features for
$t\bar{t}$ production.
\begin{description}
\item{(i)} The light quarks are almost flat in the extra
dimension to ensure a small coupling to the gauge KK modes. The LH (RH)
light quarks have a slight preference for localization towards the IR
(UV) brane that naturally makes $g_L \approx - g_R \approx
0.2-0.3 g$. Thus the coupling is naturally small, almost purely axial
and negative. 

\item{(ii)} Both components of the top (LH and RH) are
localized towards the IR brane. The localization is stronger for the
RH component, resulting in large couplings to the KK gluon that are
neither purely vector or purely axial, but with a positive and sizable
axial component.

\item{(iii)} The massive KK excitations of the EW gauge
bosons unitarize $WW$ scattering. This forces the gauge
resonances (including the gluon) to have a mass below 1 TeV.
\end{description}
These features imply that Higgsless models naturally realize the light
axigluon that we discussed in the previous section. Also, the first
two points guarantee that the axial couplings of the ligth quarks and
the top have opposite signs, thus giving a positive contribution to
$A^{t\bar{t}}$ as observed at the Tevatron. The relatively large
axial coupling of the top and light KK gluon mass make it possible to
generate a sizable asymmetry without the need of large axial couplings
for the light quarks. 
We find remarkable that all these features are
entirely imposed by constraints from EW precision data and have
nothing to do with top data.

\begin{figure}[h]
\begin{center}
\begin{tabular}{ccc}
\includegraphics[width=0.48\linewidth]{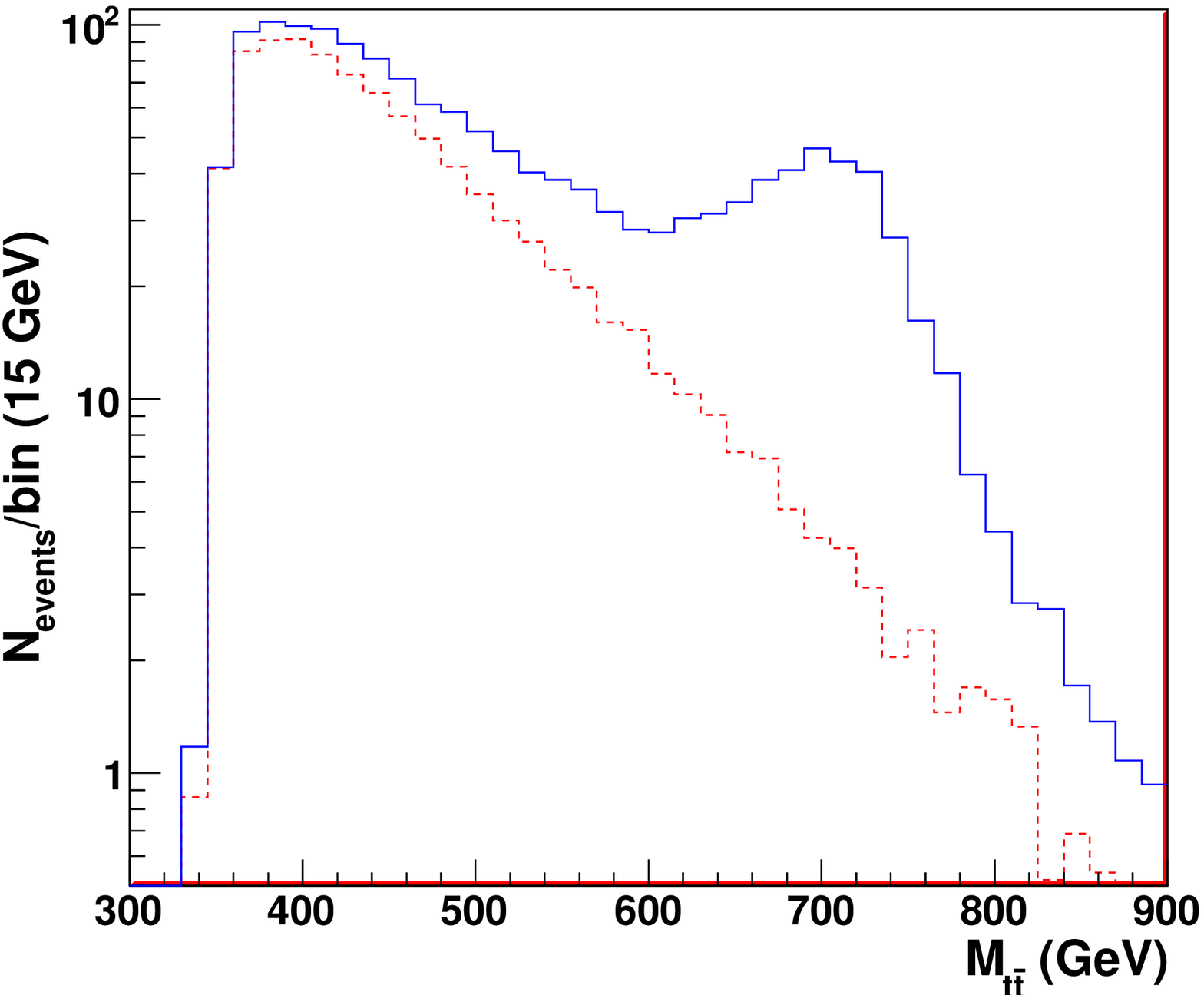} & 
\includegraphics[width=0.48\linewidth]{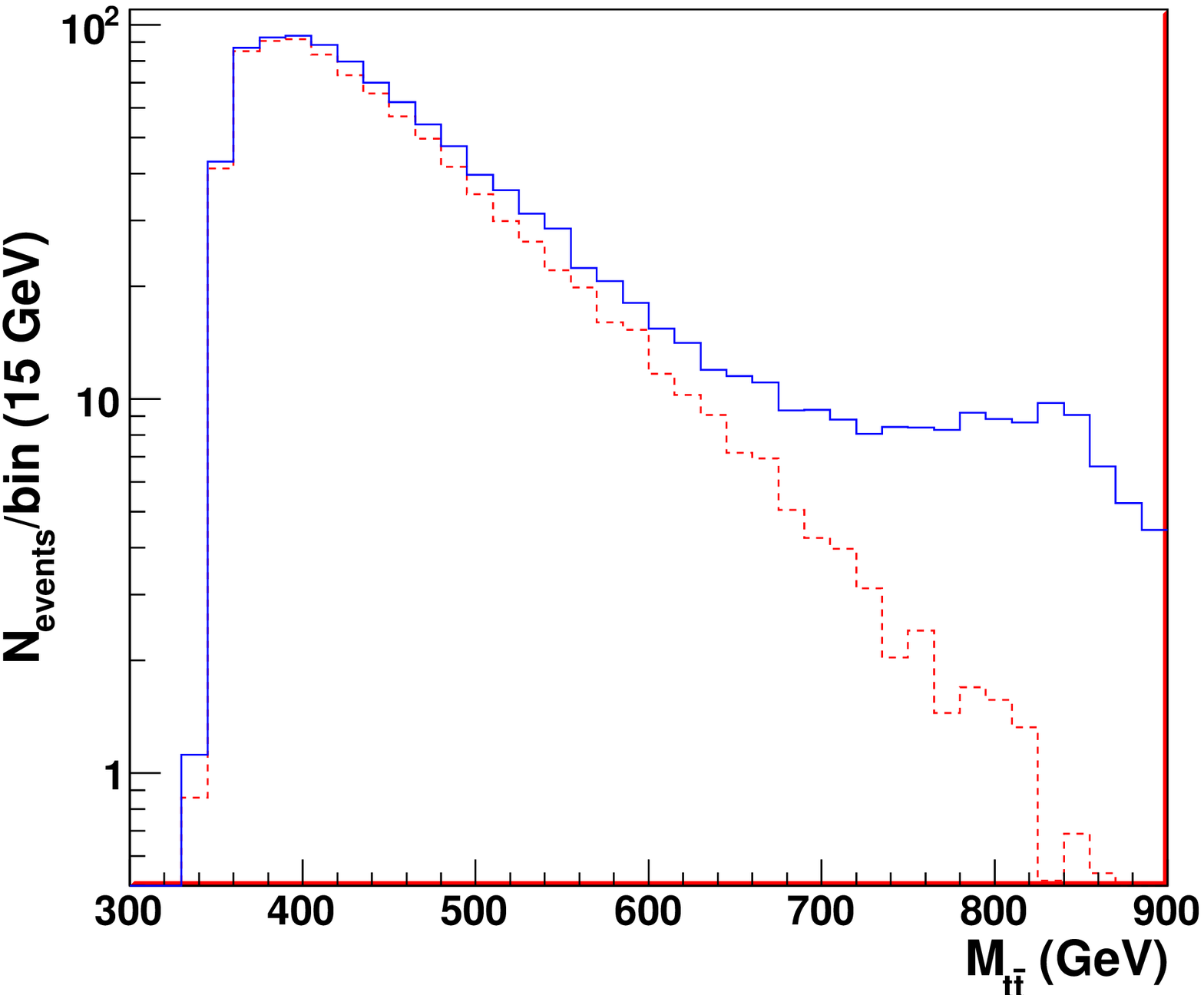} 
\end{tabular}
\end{center}
\caption{$M_{t\bar t}$ distribution at the Tevatron for Higgsless
  models. Left panel: original Higgsless model of
  Eq. (\ref{H:original}); right panel: modified Higgsless
model Eq. (\ref{H:modified}). 
In both cases the contribution in the Higgsless models
is shown in solid while the SM only contribution is shown in
dashed. We have considered a luminosity of 5.3 fb$^{-1}$.
}
\label{Higgsless:plots}
\end{figure}

The original model proposed in~\cite{Cacciapaglia:2006gp}
corresponds to a first gluon excitation with a mass $M_G=714$ GeV and 
the following couplings
\beqa
g_R^q&=&g^b_R=-0.31\,g, \qquad g_L^q=+0.17\,g, \nonumber\\
g_R^t&=&+2.27\,g, \qquad \qquad \,\, g_L^t=g^b_L=+1.93\,g, \label{H:original}
\eeqa
resulting in a  total width $\Gamma_G=0.13\,M_G$. In
Fig.~\ref{Higgsless:plots}--\textit{left} 
we plot the
invariant mass distribution for this model using again
the luminosity (5.3 fb$^{-1}$) and the cuts described in~\cite{AFB3}. 
The total number of $t\bar t$ pairs is almost a 60\% higher
than in the SM. In addition, the 275 events
between 650 and 750 GeV form a clear peak that 
should have been observed in the analysis of the Tevatron 
data.
\\
We show that with a minimal variation this model, while still
consistent with EW data, improves the agreement with Tevatron data.
The KK gluon mass is increased to 850 GeV by slightly
changing the value of the IR scale $1/R^\prime=340$ GeV (the
corresponding value of the UV scale is $1/R\approx 2.9\times 10^{10}$
GeV, see appendix for the details). 
We also optimize the localization of the different
quarks (while still being consistent with EW precision tests) 
so that the new couplings are given by
\beqa
g_R^q&=&g^b_R=-0.25\,g, \qquad g_L^q=+0.20\,g, \nonumber\\
g_R^t&=&+4.00\,g, \qquad \qquad \,\, g_L^t=g^b_L=+1.00\,g,
\label{H:modified}
\eeqa
With this choice of parameters, 
the resonance has a width 
$\Gamma_G=0.17 M_G$.
\begin{figure}[h]
\begin{center}
\begin{tabular}{ccc}
\includegraphics[width=0.48\linewidth]{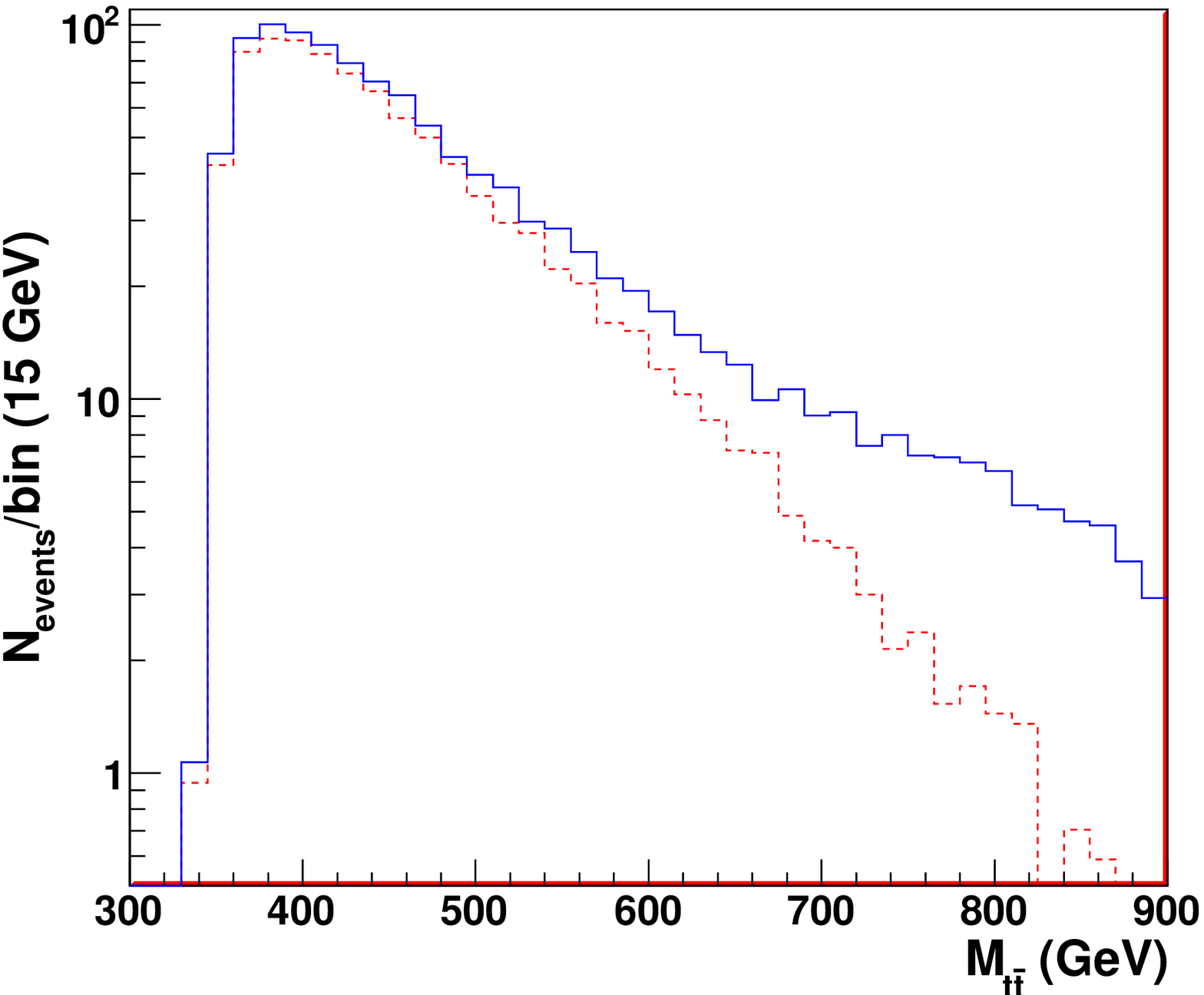} & 
\includegraphics[width=0.48\linewidth]{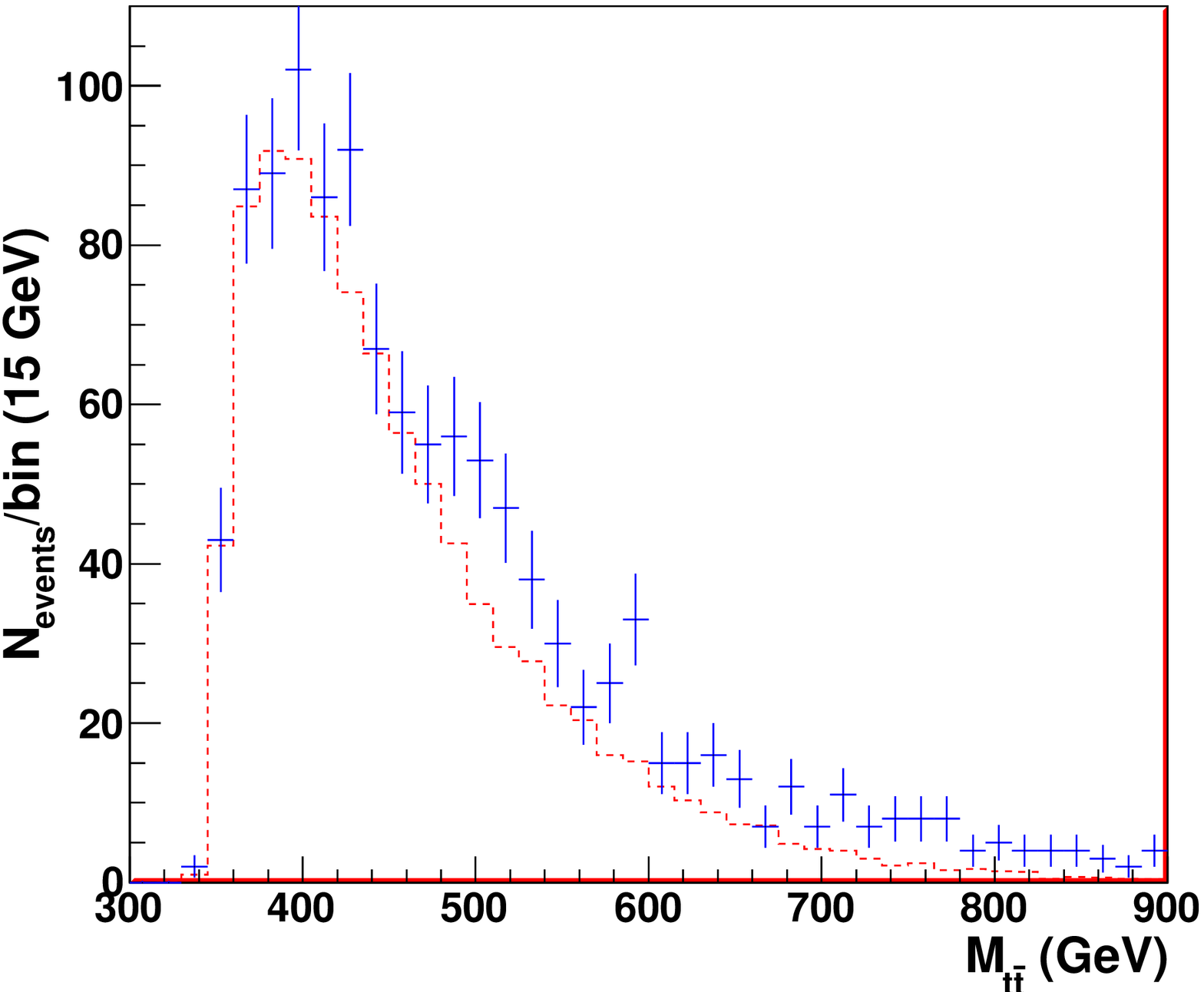} 
\end{tabular}
\end{center}
\caption{$M_{t\bar t}$ distribution at the Tevatron 
in the SM (dashes) and in the modified Higgsless model (solid) 
for a luminosity of 5.3 fb$^{-1}$.
On the right we plot a particular Montecarlo simulation.
The errors shown are statistical only.}
\label{Higgsless:motivated}
\end{figure}
We show in Fig.~\ref{Higgsless:plots}--\textit{right} 
the invariant-mass distribution of the
1113
$t\bar t$ pairs that survive the cuts. 
At $M_{t\bar t}<600$ GeV the model gives a 
8\% 
excess respect to the
SM value, whereas at higher invariant masses we obtain 
197 events versus 80 within the SM. 
This excess, together with the change in slope makes it likely that
the model should have been seen in the Tevatron data, although only a
detailed statistical analysis could state the confidence of the
exclusion. Nevertheless it is clear that the slightly higher mass, the
reduction in the vector
component of the light-quark couplings, the enhancement of the top
couplings and the increased width all go in the correct direction to
hide the KK gluon in the invariant-mass distributions while increasing
the agreement in the forward-backward asymmetry. The asymmetry for the
original Higgsless model is very small whereas we find for the
modified model
\begin{equation}
A_{\not H\mbox{ mod}}^{t\bar t} \approx \left\{
\begin{array}{l l} 
\displaystyle 0.04
& M_{t\bar t}<450\;{\rm GeV}\,; \\
0.16
& M_{t\bar t}>450\;{\rm GeV}\,.
\end{array} \right. 
\end{equation}
Adding the standard NLO contribution, 
of order $A_{NLO}^{t\bar t} \approx 0.09$ for
$M_{t\bar{t}}>450$ GeV, we obtain in the modified model a total
asymmetry less than $2\sigma$ away from the measured value.

The model we have just presented improves the agreement with the
observed asymmetry, although still at the price of making the model
likely visible in Tevatron data on the $t\bar{t}$ invariant-mass
distribution. 
The crucial point is that these models provide in a natural way (all
features are enforced by EW data, completely unrelated to the top
physics we are discussing) a framework
that realizes a light axigluon with the right couplings. Small
variations of the model can easily further improve the agreement with the
observed asymmetry without conflict with current data on the
invariant-mass distribution. In particular, it is clear that making the RH top
coupling a bit larger will increase the asymmetry and the width of
the gluon resonance, thus supressing the peak structure in the tail of the
invariant mass distribution.

As an example,
we have taken the following values of the couplings, with the same KK
gluon mass,
\beqa
g_R^q&=&g_R^b=-0.25\,g, \qquad g_L^q=+0.20\,g, \nonumber\\
g_R^t&=&+6.00\,g, \qquad \qquad \, g_L^t=g^b_L=+0.20\,g.
\label{Higgsless:motivated:eq}
\eeqa
resulting in a width $\Gamma_G=0.32 M_G$.
This model is very similar to model A in the introduction. We show in
Fig.~\ref{Higgsless:motivated} the $t\bar{t}$ invariant mass
distribution after cuts for the model and the SM contribution (left
panel) and a particular MonteCarlo simulation with the collected
luminosity to show that the differences are not statistically
significant (right panel). 
The asymmetry is increased in this case to
\beq
A_{\not H}^{t\bar t} \approx \left\{
\begin{array}{l l} 
\displaystyle 0.07
& M_{t\bar t}<450\;{\rm GeV}\,; \\
0.23
& M_{t\bar t}>450\;{\rm GeV}\,.
\end{array} \right. 
\eeq
leaving the total asymmetry just $1.4 \sigma$ below the observed
value.  
Just like for model A in the introduction, 
the cross section for $M_{t\bar{t}}<600$ GeV is a bit above the SM
expectation ($8\%$). This fact (that could influence the normalization
of the experimental data) together with the absence of any 
feature (peak) along the
tail would make the model difficult to see at the Tevatron.

\section{Invariant mass distribution at the LHC \label{LHC}}

We have seen that warped Higgsless models provide a 
framework with all the required ingredients to explain the observed
$A^{t\bar{t}}$ without conflicting with Tevatron data on the total cross
section. Even with the low integrated luminosity collected by LHC in the
2010 run, analyses of dijet and $t\bar{t}$ data are beginning to probe
the parameter space of many models proposed to explain the Tevatron
asymmetry. In this section we show that the Higgsless motivated model
of Eq. (\ref{Higgsless:motivated:eq})
cannot be seen with the current luminosity 
but should be either discovered or excluded
with 2011 data.
Also the small couplings to light quarks and
vanishing couplings to SM gluons make the model invisible in the dijet
sample.
 
\begin{figure}[h]
\begin{center}
\begin{tabular}{ccc}
\includegraphics[width=0.48\linewidth]{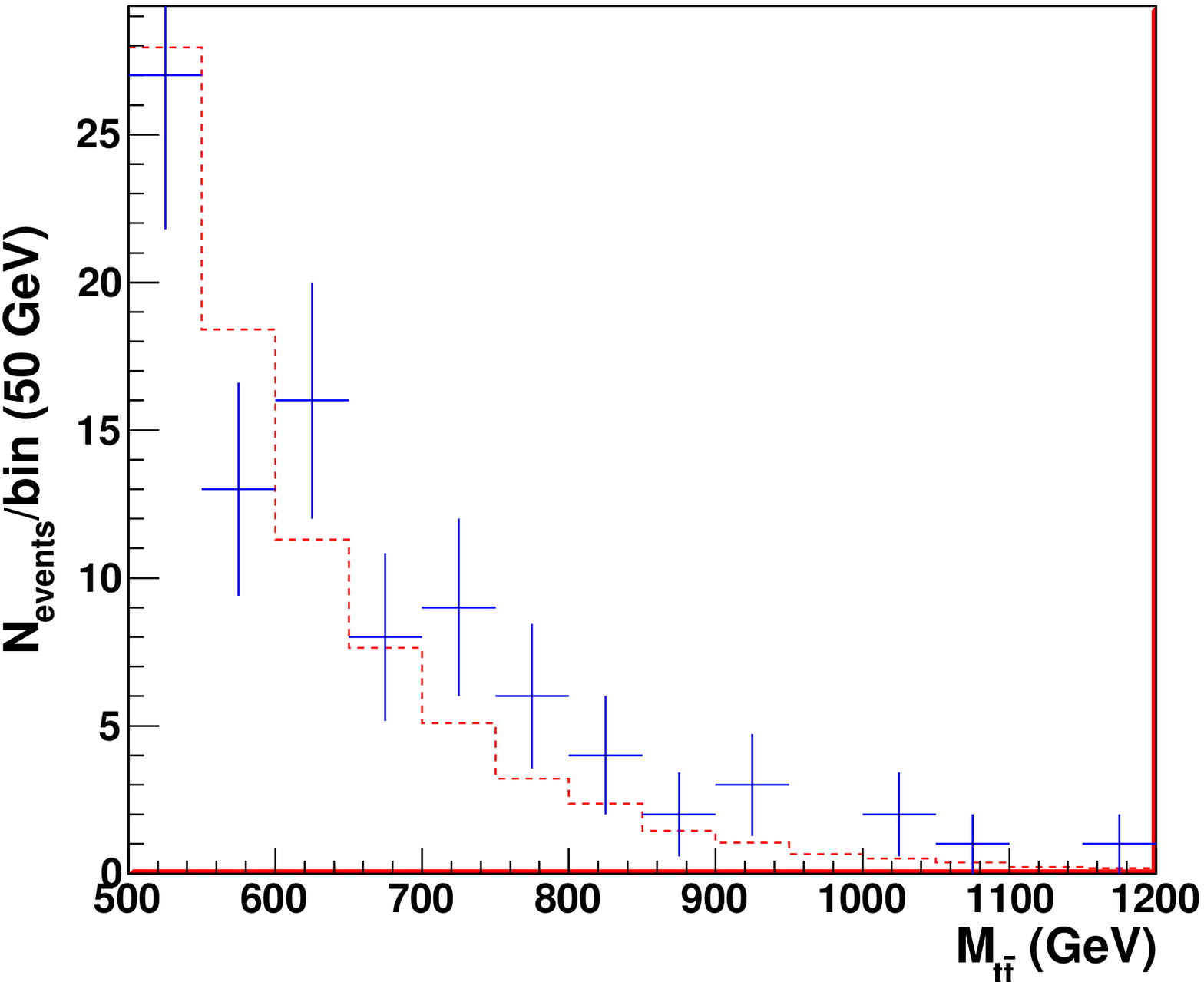} & 
\includegraphics[width=0.48\linewidth]{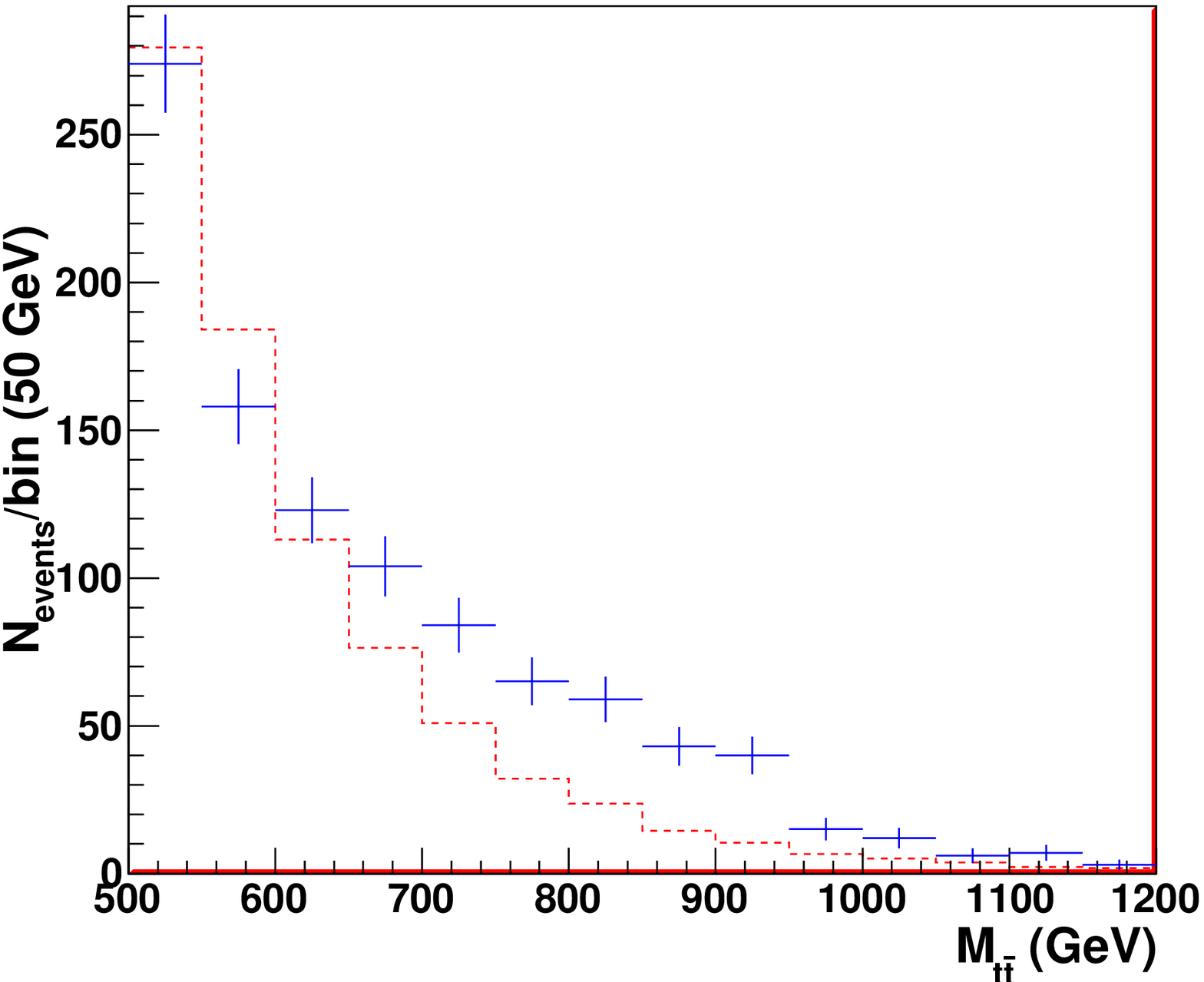} 
\end{tabular}
\end{center}
\caption{$M_{t\bar t}$ distribution at the LHC
in the SM (dashes) and in the Higgsless motivated model of
Eq. (\ref{Higgsless:motivated:eq}) (solid) 
for a luminosity of 36 pb$^{-1}$ (left) and 360 pb$^{-1}$ (right). We
show particular Montecarlo simulations corresponding to the simulated
luminosity. Errors shown are statistical only.}
\label{LHCttbar:fig}
\end{figure}
We have considered a luminosity of 36 pb$^{-1}$ at 7 TeV. In
Fig.~\ref{LHCttbar:fig}--left we 
plot the number of events after cuts per 50 GeV bin, as described
in~\cite{CMS:ttbar}, for a particular Montecarlo simulation. 
We obtain 24 events in the 700-1000 GeV interval for our model versus
14 events for the SM. 
It is apparent that the low number of events makes {\it invisible} the
peak around $M_{t\bar t}=850$ GeV. 
Increasing the luminosity by
a factor of 10 (right panel of the figure) 
we obtain 306 events in the 700-1000 GeV interval 
(versus just 138 in the SM). The excess in the signal should be enough to 
provide evidence for this type of gluon excitation.

\begin{figure}
\begin{center}
\includegraphics[width=0.7\linewidth]{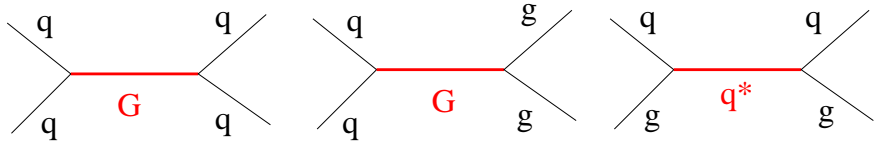} 
\end{center}
\caption{Dijet processes through KK gluons ($G$) and quarks ($q^*$). 
Possible $t$ and $u$ channel contributions are not explicitly shown. The first
amplitude is suppressed by the couplings to the light quarks, whereas
the other two vanish due to the orthogonality of the wave functions.}
\label{fig3}
\end{figure}

As for dijet signals, the particular features of the model under
consideration make it completely invisible.
We show in Fig.~\ref{fig3} the leading dijet production mechanism
through massive gluon or quark resonances. The first diagram has the
suppression of the couplings to the light quarks giving a branching
ratio $BR(G\to q\bar{q})\approx 2 \%$. The ATLAS
analysis~\cite{Aad:2011aj} is sensitive to a cross section of the
order of a $10\%$ of the one expected for an axigluon of mass $850$
GeV with $g_A=g$ to the light quarks, well above the suppression in
our model. To confirm this expectation we have simulated dijet events
at the LHC and found that less than around $0.5\%$ of the events pass
the cuts in the analysis of~\cite{Aad:2011aj} (the extra suppression
is due to the dominant gluon initiated processes that remain unchanged
in our model). 
The second and third
diagrams of Fig.~\ref{fig3} exactly vanish in our model due to
the orthogonality of the zero and the massive wave functions.

\section{Summary and discussion \label{conclusions}}

A strong forward-backward asymmetry may seem a very unexpected
feature in the usual scenarios for physics beyond the 
standard model. We have argued that realistic Higgsless models with
warped extra dimensions naturally provide a general framework to 
generate such an asymmetry.
In these models one expects massive gluon excitations 
strongly coupled to the top quark, with much smaller 
couplings to light quarks and gluons (as required by
dijet searches at the LHC~\cite{Aad:2011aj}). In addition, the
vector coupling of the light quarks may be weaker than
the axial-vector one ($g_V^q\ll g_A^q$), which suppresses
anomalies in the $t\bar t$ invariant mass distribution
while introducing forward-backward asymmetry (note that the couplings
of the top quark do not need to be mostly axial to cancel the largest
contributions to the total cross section).
It is remarkable that all these features are imposed on Higgsless
models by EW physics rather than top physics. In particular, EW
constraints force the new physics contribution to 
$A_{t\bar{t}}$ to be positive, as experimentally
observed. 

We have shown that, while
the original Higgsless model is excluded by Tevatron top data, a
slight modification can increase the agreement with
the $t\bar{t}$ asymmetry while making it (barely) consistent with the
invariant-mass distribution.
This has motivated us to propose a Higgsless inspired model that is
compatible with Tevatron data.

The model consists of a massive gluon of $\approx 850$ GeV
with mostly axial-vector couplings to the light quarks and both
vector and axial-vector couplings to the top quark. We show
that the Tevatron does not have enough energy to {\it see} a peak
at $M_{t\bar t}\approx M_G$, whereas the change in the slope 
of the  $M_{t\bar t}$ distribution  at  $M_{t\bar t}\approx 650$ GeV
is of little statistical 
significance. We have also shown that the LHC has enough energy to
reach the resonance but not yet enough integrated luminosity. We have
seen that about 10 times more luminosity than
the current 36 pb$^{-1}$ should be enough to probe the model. Finally,
the suppressed coupling to light quarks makes the new resonance
virtually invisible in dijet data.

There are some aspects of Higgsless models that we have not fully
explored here and deserve further investigation. 
For instance, the optimal values of the
couplings for the third generation quarks in the minimal realistic
set-up presented in~\cite{Cacciapaglia:2006gp,Csaki:2009bb}
bring the disagreement with the observed asymmetry down to $\approx 2
\sigma$, but it seems difficult to improve this results in the context of
these minimal models. Furthermore, even with such modifications, there
is still a small peak that might be observable in the Tevatron data. 
It would be interesting to see if simple
modifications, for instance in the gravitational background, could
allow for a larger axial top coupling to improve the agreement with
$A^{t\bar{t}}$ and the invariant mass distribution. 
We have shown that such a modification improves
the level of agreement with current data on the total cross section
and the asymmetry at the Tevatron while making it invisible at 
the LHC (both in $t\bar{t}$ and dijet data). Also, we have not
included the effect of fermion KK excitations. Some of them can be
relatively light and influence the collider implications of the KK
gluon. It would be interesting to see in which direction these
modifications go and the interplay between KK gluon searches and
fermion KK searches at the LHC (see~\cite{Carena:2007tn} 
for an example of these effects). In principle, the new quarks would
increase the width of the resonance and decrease the branching ratio
into top pairs thus making the peak even more difficult to detect at
the Tevatron.

\appendix

\section{Details of the Higgsless Model}

Let us briefly review the most relevant features of 
a realistic Higgsless Model with flavor protection. Full details can
be found in~\cite{Cacciapaglia:2006gp,Csaki:2009bb}. The model lives
in 5D with a metric 
\begin{equation}
ds^2=\left(\frac{R}{z}\right)^2 [dx^2 - dz^2],
\end{equation}
where the extra dimension is bounded $R\leq z \leq R^\prime$ by the UV
and IR branes, respectively. The bulk gauge symmetry is $SU(3)_C
\times SU(2)_L\times
SU(2)_R\times U(1)_X$, broken by boundary conditions to $SU(3)_C
\times U(1)_Q$. We focus on the quark sector. 
The first two generations live in $(2,1)$ multiplets of $SU(2)_L
\times SU(2)_R$ for the LH
components and in $(1,2)$ for the RH ones (they are all color triplets
and have $Q_X=\frac{1}{6}$). The flavor
symmetry forces the localization of the two $(2,1)$ multiplets to be
the same and similarly for the $(1,2)$ multiplets. The third
generation is in an almost custodially protected
representation
\begin{eqnarray}
\Psi_l= \begin{pmatrix}
t_l[+,+]&X_l[-,+]\\b_l[+,+]&T_l[-,+]\end{pmatrix}&\sim&
(2,2),\quad \Psi_r=\begin{pmatrix} 
X_r[+,-]\\T_r[+,-]\\b_r[-,-]
\end{pmatrix}\sim (1,3),\nonumber\\
&&t_r[-,-]\sim (1,1).
\end{eqnarray}
In this case all multiplets have $Q_X=\frac{2}{3}$ and 
the left and right columns of the bidoublet correspond to fields with
$T_R^3=\mp 1/2$ while the upper and lower components have $T_L^3=\pm
1/2$. The signs in square brackets are 
a shorthand for the boundary conditions in the absence of localized
brane terms. A Dirichlet boundary condition for the right-handed (RH)
component is denoted by $[+]$, whereas $[-]$ denotes a Dirichlet
boundary condition for  
the left-handed (LH) chirality. The
first sign corresponds to the boundary condition at the 
UV brane and the second one at the IR brane. 

These boundary conditions are changed on the IR brane due to the
presence of the following localized mass terms 
\begin{eqnarray}
-\mathcal{S}_{IR}&=&\int\mathrm{d}x^4\int_{R}^{R^{\prime}}
\mathrm{d}z\left(\frac{R}{z}\right)^4
\delta(z-R^{\prime})\left\{M_3\left[\frac{1}{\sqrt{2}}
\psi_{T_r}\left(\chi_{t_l}+\chi_{T_l}\right)+\psi_{b_r}\chi_{b_l}
+\psi_{X_r}\chi_{X_l}\right] 
\right.\nonumber\\
&+&
\left.\frac{M_1}{\sqrt{2}}\psi_{t_r}
\left(\chi_{t_l}-\chi_{T_l}\right)\right\}+\mathrm{h.c.}~, 
\end{eqnarray}
where we denote with $\chi_{\Psi}$ ($\bar{\psi}_{\Psi}$) the LH (RH)
component of field $\Psi$.
These localized masses 
allow us to give a mass to the third generation of quarks and,
as explained in \cite{Cacciapaglia:2006gp}, keep $Zb_L\bar{b}_L$
corrections under control.

To check EWPT we canonically normalize the SM gauge fields and obtain,
for a fixed value of $R^{\prime}$,  the parameters $R,~g_{5L}=g_{5R}$
and $g_{5X}$ in terms of the measured values of $M_W,M_Z$ and the
electromagnetic coupling $e(M_Z)$. We take  the PDG's values
\cite{Nakamura:2010zzi}, $M_W=80.399~\GeV, ~M_Z=91.1876~\GeV$ and
$e(M_Z)=\sqrt{4\pi/128}$. As the first KK gluon mass is given roughly
by $m_{G}R^{\prime}\sim 2.5$ we choose $R^{\prime}=2.5/0.850
~\TeV^{-1}$ obtaining $m_{G}=0.848~\TeV$. 
We neglect the mass of the first two generations and 
work in the zero mode approximation. In that case, choosing
$c_L=0.466$ and $c_R=-0.65$ we obtain the following shifts for the
$Zd\bar{d}$ vertex 
\begin{eqnarray}
\delta g_{d_R}^Z/g_{d_R}^{Z\,SM}\sim -0.27\%,\qquad \delta
g_{d_R}^Z/g_{d_R}^{Z\,SM}\sim -0.30\%, 
\end{eqnarray}
and similar deviations in the up sector.
We consider these values reasonably compatible with EW precision
data.~\footnote{It should be noted that higher dimensional operators
  could give a non-negligible contribution as could in general be needed to
  compensate for calculable one loop corrections in these
  models~\cite{oneloop}.}  
For these values of the bulk masses,
the couplings to the KK gluon G are the following  
\begin{eqnarray}
g_R^q=-0.26g,\qquad g_L^q=0.19g.
\end{eqnarray}

Regarding the third generation, for each value of the bulk mass parameters
$c_{\Psi_L}, c_{\Psi_R}$ and $c_{t_t}$, we fix $M_1$ and $M_3$ to
reproduce the 
top and bottom masses, $m_t=170 ~\GeV$ and $m_b=4~\GeV$.  
We find that the top mass cannot be generated, for any value of $M_1$
and $M_3$, unless $c_{\Psi_l} \lesssim 0.35$. Thus, we take $c_{\Psi_l}=0.35$. 
We also choose $c_{\Psi_r}=-0.677$ so that the
corrections to the $Zb\bar{b}$ vertex that are allowed by EWPT
\cite{Choudhury:2001hs}: 
\begin{eqnarray}
\delta g_{b_L}^Z/g_{b_L}^{Z\,SM}\sim -0.08\%,\qquad \delta
g_{b_R}^Z/g_{b_R}^{Z\,SM}\sim 2.5\%. 
\end{eqnarray}
Finally, the $Gt_R\bar{t}_R$ coupling is maximized for large values of
$c_{t_R}$ although it saturates for $c_{t_R}\gtrsim 1$. We choose $c_{t_R}=1.6$.  
With these values of the bulk masses we obtain the following couplings
to the first KK gluon $G$: 
\begin{eqnarray}
g_L^{t}&=&+1.06g\qquad g_R^{t}=3.95g\\
g_L^{b}&=&+1.39g\qquad g_R^{b}=-0.28g.
\end{eqnarray}

\section*{Acknowledgments}
We would like to thank J.A. Aguilar-Saavedra and M. P\'erez-Victoria
and S. Westhoff
for useful discussions.
This work has been partially supported by
MICINN of Spain (FPA2006-05294, FPA2010-16802, FPA2010-17915,
Consolider-Ingenio 
{\bf Multidark} CSD2009-00064 and Ram\'on y Cajal Program)  
and by Junta de Andaluc\'{\i}a
(FQM 101, FQM 3048 and FQM 6552).


\begin{thebibliography}{99}

%\cite{AFB1}
\bibitem{AFB1}
  V.~M.~Abazov {\it et al.} [ D0 Collaboration ],
  %``First measurement of the forward-backward charge asymmetry in top quark pair production,''
  Phys.\ Rev.\ Lett.\  {\bf 100 } (2008)  142002.
  [arXiv:0712.0851 [hep-ex]].
%\cite{Aaltonen:2008hc}
\bibitem{AFB2}
  T.~Aaltonen {\it et al.} [ CDF Collaboration ],
  %``Forward-Backward Asymmetry in Top Quark Production in $p\bar{p}$ Collisions at $sqrt{s}=1.96$ TeV,''
  Phys.\ Rev.\ Lett.\  {\bf 101 } (2008)  202001.
  [arXiv:0806.2472 [hep-ex]].
%\cite{Aaltonen:2011kc}
\bibitem{AFB3}
  T.~Aaltonen {\it et al.} [ CDF Collaboration ],
  %``Evidence for a Mass Dependent Forward-Backward Asymmetry in Top Quark Pair Production,''
  %Submitted to: Phys.Rev.D.
  [arXiv:1101.0034 [hep-ex]].




%%%%%%%%%%% AXIGLUONS

%\cite{Ferrario:2009bz}
\bibitem{Ferrario:2009bz}
  P.~Ferrario, G.~Rodrigo,
  %``Constraining heavy colored resonances from top-antitop quark events,''
  Phys.\ Rev.\  {\bf D80 } (2009)  051701.
  [arXiv:0906.5541 [hep-ph]].

%\cite{Antunano:2007da}
\bibitem{Antunano:2007da}
  O.~Antunano, J.~H.~Kuhn, G.~Rodrigo,
  %``Top quarks, axigluons and charge asymmetries at hadron colliders,''
  Phys.\ Rev.\  {\bf D77 } (2008)  014003.
  [arXiv:0709.1652 [hep-ph]];
%\cite{Djouadi:2009nb}
%\bibitem{Djouadi:2009nb}
  A.~Djouadi, G.~Moreau, F.~Richard, R.~K.~Singh,
  %``The Forward-backward asymmetry of top quark production at the Tevatron in warped extra dimensional models,''
  Phys.\ Rev.\  {\bf D82 } (2010)  071702.
  [arXiv:0906.0604 [hep-ph]];
%\cite{Frampton:2009rk}
%\bibitem{Frampton:2009rk}
  P.~H.~Frampton, J.~Shu, K.~Wang,
  %``Axigluon as Possible Explanation for p anti-p ---> t anti-t Forward-Backward Asymmetry,''
  Phys.\ Lett.\  {\bf B683}, 294-297 (2010).
  [arXiv:0911.2955 [hep-ph]];
%\cite{Cao:2010zb}
%\bibitem{Cao:2010zb}
  Q.~-H.~Cao, D.~McKeen, J.~L.~Rosner, G.~Shaughnessy, C.~E.~M.~Wagner,
  %``Forward-Backward Asymmetry of Top Quark Pair Production,''
  Phys.\ Rev.\  {\bf D81 } (2010)  114004.
  [arXiv:1003.3461 [hep-ph]];
%\cite{Choudhury:2010cd}
%\bibitem{Choudhury:2010cd}
  D.~Choudhury, R.~M.~Godbole, S.~D.~Rindani, P.~Saha,
  %``Top polarization, forward-backward asymmetry and new physics,''  
  [arXiv:1012.4750 [hep-ph]];
%\cite{Bai:2011ed}
%\bibitem{Bai:2011ed}
  Y.~Bai, J.~L.~Hewett, J.~Kaplan, T.~G.~Rizzo,
  %``LHC Predictions from a Tevatron Anomaly in the Top Quark Forward-Backward Asymmetry,''
  JHEP {\bf 1103 } (2011)  003.
  [arXiv:1101.5203 [hep-ph]];
%\cite{Gresham:2011pa}
%\bibitem{Gresham:2011pa}
  M.~I.~Gresham, I.~-W.~Kim, K.~M.~Zurek,
  %``On Models of New Physics for the Tevatron Top $A_{FB}$,''  
  [arXiv:1103.3501 [hep-ph]].

%%%%%%%%%%% MODEL INDEPENDENT
%\cite{Degrande:2010kt}
\bibitem{Degrande:2010kt}
  C.~Degrande, J.~-M.~Gerard, C.~Grojean, F.~Maltoni, G.~Servant,
  %``Non-resonant New Physics in Top Pair Production at Hadron Colliders,''
  JHEP {\bf 1103 } (2011)  125.
  [arXiv:1010.6304 [hep-ph]];
%\cite{Delaunay:2011gv}
%\bibitem{Delaunay:2011gv}
  C.~Delaunay, O.~Gedalia, Y.~Hochberg, G.~Perez, Y.~Soreq,
  %``Implications of the CDF $t \bar{t}$ Forward-Backward Asymmetry for Hard Top Physics,''  
  [arXiv:1103.2297 [hep-ph]];
%\cite{AguilarSaavedra:2011vw}
%\bibitem{AguilarSaavedra:2011vw}
  J.~A.~Aguilar-Saavedra, M.~Perez-Victoria,
  %``Probing the Tevatron t tbar asymmetry at LHC,''
  [arXiv:1103.2765 [hep-ph]];
%\cite{AguilarSaavedra:2011zy}
%\bibitem{AguilarSaavedra:2011zy}
%  J.~A.~Aguilar-Saavedra, M.~Perez-Victoria,
  %``No like-sign tops at Tevatron: Constraints on extended models and implications for the t tbar asymmetry,'' 
  [arXiv:1104.1385 [hep-ph]];
%\cite{Degrande:2011rt}
%\bibitem{Degrande:2011rt}
  C.~Degrande, J.~-M.~Gerard, C.~Grojean, F.~Maltoni, G.~Servant,
  %``An effective approach to same sign top pair production at the LHC and the forward-backward asymmetry at the Tevatron,'' 
  [arXiv:1104.1798 [hep-ph]].


%\cite{Burdman:2010gr}
\bibitem{Burdman:2010gr}
  G.~Burdman, L.~de Lima, R.~D.~Matheus,
  %``New Strongly Coupled Sector at the Tevatron and the LHC,''
  Phys.\ Rev.\  {\bf D83}, 035012 (2011).
  [arXiv:1011.6380 [hep-ph]].

%%%%%%%%%%%%%%%%%%%



%\cite{Csaki:2003dt}
\bibitem{Csaki:2003dt}
  C.~Csaki, C.~Grojean, H.~Murayama, L.~Pilo, J.~Terning,
  %``Gauge theories on an interval: Unitarity without a Higgs,''
  Phys.\ Rev.\  {\bf D69}, 055006 (2004).
  [hep-ph/0305237];
%\cite{Csaki:2003zu}
%\bibitem{Csaki:2003zu}
  C.~Csaki, C.~Grojean, L.~Pilo, J.~Terning,
  %``Towards a realistic model of Higgsless electroweak symmetry breaking,''
  Phys.\ Rev.\ Lett.\  {\bf 92}, 101802 (2004).
  [hep-ph/0308038].

%\cite{Cacciapaglia:2006gp}
\bibitem{Cacciapaglia:2006gp}
  G.~Cacciapaglia, C.~Csaki, G.~Marandella, J.~Terning,
  %``A New custodian for a realistic Higgsless model,''
  Phys.\ Rev.\  {\bf D75 } (2007)  015003.
  [hep-ph/0607146].

%\cite{Alwall:2007st}
\bibitem{Alwall:2007st}
  J.~Alwall {\it et al.},
  %``MadGraph/MadEvent v4: The New Web Generation,''
  JHEP {\bf 0709} (2007) 028
  [arXiv:0706.2334 [hep-ph]].
  %%CITATION = JHEPA,0709,028;%%

%\cite{Sjostrand:2006za}
\bibitem{pythia}
  T.~Sjostrand, S.~Mrenna and P.~Z.~Skands,
  %``PYTHIA 6.4 Physics and Manual,''
  JHEP {\bf 0605} (2006) 026
  [arXiv:hep-ph/0603175].
  %%CITATION = JHEPA,0605,026;%%

\bibitem{PGS4} PGS4
http://www.physics.ucdavis.edu/$\sim$
conway/research/software/pgs/pgs4-general.htm 


%\cite{Campbell:1999ah}
\bibitem{Campbell:1999ah}
  J.~M.~Campbell, R.~K.~Ellis,
  %``An Update on vector boson pair production at hadron colliders,''
  Phys.\ Rev.\  {\bf D60}, 113006 (1999).
  [hep-ph/9905386].


%\cite{Bauer:2010iq}
\bibitem{Bauer:2010iq}
  M.~Bauer, F.~Goertz, U.~Haisch, T.~Pfoh, S.~Westhoff,
  %``Top-Quark Forward-Backward Asymmetry in Randall-Sundrum Models Beyond the Leading Order,''
  JHEP {\bf 1011 } (2010)  039.
  [arXiv:1008.0742 [hep-ph]].



%\cite{Csaki:2009bb}
\bibitem{Csaki:2009bb}
  C.~Csaki, D.~Curtin,
  %``A Flavor Protection for Warped Higgsless Models,''
  Phys.\ Rev.\  {\bf D80 } (2009)  015027.
  [arXiv:0904.2137 [hep-ph]].


%\cite{Chivukula:2010fk}
\bibitem{Chivukula:2010fk}
  R.~S.~Chivukula, E.~H.~Simmons, C.~-P.~Yuan,
  %``Axigluons cannot explain the observed top quark forward-backward asymmetry,''
  Phys.\ Rev.\  {\bf D82 } (2010)  094009.
  [arXiv:1007.0260 [hep-ph]].

\bibitem{CMS:ttbar}
CMS Collaboration, public note CMS PAS TOP-10-007.

%\cite{Aad:2011aj}
\bibitem{Aad:2011aj}
  G.~Aad {\it et al.} [ ATLAS Collaboration ],
  %``Search for New Physics in Dijet Mass and Angular Distributions in pp Collisions at $\sqrt{s} = 7$ TeV Measured with the ATLAS Detector,''  
  [arXiv:1103.3864 [hep-ex]].

%\cite{Carena:2007tn}
\bibitem{Carena:2007tn}
  M.~Carena, A.~D.~Medina, B.~Panes, N.~R.~Shah, C.~E.~M.~Wagner,
  %``Collider phenomenology of gauge-Higgs unification scenarios in warped extra dimensions,''
  Phys.\ Rev.\  {\bf D77 } (2008)  076003.
  [arXiv:0712.0095 [hep-ph]].


%\cite{Nakamura:2010zzi}
\bibitem{Nakamura:2010zzi}
  K.~Nakamura {\it et al.} [ Particle Data Group Collaboration ],
  %``Review of particle physics,''
  J.\ Phys.\ G {\bf G37 } (2010)  075021.

\bibitem{oneloop}
%\cite{Matsuzaki:2006wn}
%\bibitem{Matsuzaki:2006wn}
  S.~Matsuzaki, R.~S.~Chivukula, E.~H.~Simmons, M.~Tanabashi,
  %``One-Loop Corrections to the S and T Parameters in a Three Site Higgsless Model,''
  Phys.\ Rev.\  {\bf D75}, 073002 (2007).
  [hep-ph/0607191];
%\cite{Sekhar Chivukula:2007ic}
%\bibitem{Sekhar Chivukula:2007ic}
  R.~S.~Chivukula, E.~H.~Simmons, S.~Matsuzaki, M.~Tanabashi,
  %``The Three Site Model at One-Loop,''
  Phys.\ Rev.\  {\bf D75}, 075012 (2007).
  [hep-ph/0702218 [HEP-PH]];
%\cite{Dawson:2008as}
%\cite{Dawson:2007yk}
%\bibitem{Dawson:2007yk}
  S.~Dawson, C.~B.~Jackson,
  %``Chiral-logarithmic Corrections to the S and T Parameters in Higgsless Models,''
  Phys.\ Rev.\  {\bf D76}, 015014 (2007).
  [hep-ph/0703299 [HEP-PH]];
%\bibitem{Dawson:2008as}
%  S.~Dawson, C.~B.~Jackson,
  %``One-loop Corrections to the S Parameter in the Four-site Model,''
  Phys.\ Rev.\  {\bf D79}, 013006 (2009).
  [arXiv:0810.5068 [hep-ph]];
%\cite{Abe:2009ni}
%\bibitem{Abe:2009ni}
  T.~Abe, R.~S.~Chivukula, N.~D.~Christensen, K.~Hsieh, S.~Matsuzaki, E.~H.~Simmons, M.~Tanabashi,
  %``Z ---> b anti-b and Chiral Currents in Higgsless Models,''
  Phys.\ Rev.\  {\bf D79}, 075016 (2009).
  [arXiv:0902.3910 [hep-ph]].


  
%\cite{Choudhury:2001hs}
\bibitem{Choudhury:2001hs}
  D.~Choudhury, T.~M.~P.~Tait, C.~E.~M.~Wagner,
  %``Beautiful mirrors and precision electroweak data,''
  Phys.\ Rev.\  {\bf D65 } (2002)  053002.
  [hep-ph/0109097].



\end{thebibliography}
\end{document}